\documentclass[a4paper,fleqn,usenatbib]{mnras}

\usepackage{newtxtext,newtxmath}

\usepackage[T1]{fontenc}
\usepackage{ae,aecompl}
\usepackage{lscape}
\usepackage{longtable}
\usepackage{hyperref}
\usepackage{wasysym} 

\newcommand{\bll}{$L_{\textrm{\scriptsize BLL}}$}
\newcommand{\fsrq}{$L_{\textrm{\scriptsize FSRQ}}$}

\usepackage{graphicx}	

\usepackage[switch]{lineno}

\title[Classifying 3FGL Unassociated Sources]{3FGLzoo. Classifying 3FGL Unassociated Fermi-LAT Gamma-ray Sources by Artificial Neural Networks}

\author[D. Salvetti et al.]{D. Salvetti$^{1}$\thanks{E-mail: salvetti@iasf-milano.inaf.it},
G. Chiaro$^{1, 2}$\thanks{Email: chiaro@pd.infn.it},
G. La Mura$^{2}$,
and D.~J. Thompson$^{3}$
\newauthor
\\
$^{1}$ INAF--Istituto di Astrofisica Spaziale e Fisica Cosmica , I-20133, Milano, Italy\\
$^{2}$ Dip. Fisica and Astronomia G. Galilei - Universit\`a di Padova, I-35131, Padova, Italy\\
$^{3}$ NASA Goddard Space Flight Center, 8800 Greenbelt Rd, Greenbelt, MD 20771, USA}

\date{Accepted XXX. Received YYY; in original form ZZZ}

\pubyear{2017}

\begin{document}


\label{firstpage}
\pagerange{\pageref{firstpage}--\pageref{lastpage}}
\maketitle

\begin{abstract}
In its first four years of operation, the {\it Fermi} Large Area Telescope (LAT) detected 3033  $\gamma$-ray emitting sources. In the {\it Fermi}-LAT Third Source Catalogue (3FGL) about 50$\%$ of the sources have no clear association with a likely $\gamma$-ray emitter. We use an artificial neural network algorithm aimed at distinguishing BL Lacs from FSRQs to investigate the source subclass of 559 3FGL unassociated sources characterised by $\gamma$-ray properties very  similar to those of Active Galactic Nuclei. Based on our method, we can classify 271 objects as BL Lac candidates, 185 as FSRQ candidates, leaving only 103 without a clear classification. we suggest a new zoo for $\gamma$-ray objects, where the percentage of sources of uncertain type drops from 52$\%$ to less than 10$\%$. The result of this study opens up new considerations on the population of the $\gamma$-ray sky, and it will facilitate the planning of significant samples for rigorous analyses and multiwavelength observational campaigns. 
\end{abstract}



\section{Introduction}

The {\it Fermi} Large Area Telescope (LAT) has provided the most comprehensive view of the $\gamma$-ray sky in the 100 MeV$-$300 GeV energy range \citep{FLAT}. The most recent catalogue of $\gamma$-ray sources detected by the LAT, the third {\it Fermi} Large Area Telescope source catalogue \citep[3FGL;][]{3FGL}, is  based on data collected in four years of operation, from 2008 August 4 (MJD 54682) to 2012 July 31 (MJD 56139)\footnote{Data are available from the Fermi Science Support Center website:\\ {\tt http://fermi.gsfc.nasa.gov/ssc/data/access/lat/4yr\_catalog/}} and contains 3033 sources. The two largest $\gamma$-ray source classes are Active Galactic Nuclei (AGN), with 1745 objects, and pulsars (PSR), with 167 objects. Out of 1745 AGN, 1144 are blazars, subdivided into 660 BL Lacertae (BLL) and 484 Flat Spectrum Radio Quasars (FSRQ). The catalogue includes also 573 blazars of uncertain type (BCU), i.e. $\gamma$-ray sources positionally coincident with an object showing distinctive broad-band blazar characteristics but lacking reliable optical spectrum measurements. In addition, 30 per cent of the 3FGL sources, 1010 objects, have not even a tentative association with a likely $\gamma$-ray-emitting object and are referred to as unassociated sources.  As a result, the nature of about half the $\gamma$-ray sources, i.e. BCU and unassociated sources, is still not completely known. Since blazars are the most numerous $\gamma$-ray source class, we expect that a large fraction of unassociated sources might belong to one of its subclasses, BLLs or FSRQs. 

Rigorous determination of whether an unassociated source is a BLL or a FSRQ requires the optical spectrum of the correct counterpart. FSRQs have strong, broad emission lines at optical wavelengths, while BLLs show at most weak emission lines, sometimes display absorption features, and can also be completely featureless \citep{abd10}. For this reason detailed optical spectral observation campaigns to identify the nature of many unassociated sources are in progress \citep[e.g.][]{lan15,mas16,alv16,mar16}. Unfortunately, optical observations are demanding and time consuming. An easy screening method to suggest the nature of a $\gamma$-ray source counterpart could be very useful for the scientific community in order to plan new focused observational campaigns and research projects. Machine-learning techniques are powerful tools for screening and ranking objects according to their \emph{predicted} classification. Recently, \citet{pablo} developed a method based on machine-learning techniques to distinguish pulsars from AGN candidates among 3FGL unassociated sources using only $\gamma$-ray data. In this work we explore the possibility of applying our Blazar Flaring Pattern (B-FlaP) algorithm, which is based on an artificial neural network technique \citep[for a full description see][]{bflap}, to provide a preliminary and reliable identification of AGN-like unassociated sources as likely BLL or FSRQ candidates.

The paper is organized as follows: in Sectn.~\ref{sec:2} we provide a brief description of the machine-learning technique we employed, in Sectn.~\ref{sec:3} we present results of the algorithm at classifying 3FGL unassociated sources and we test our predictions through optical spectral observations of a number of targets, and we discuss implications of our results in Sectn.~\ref{sec:5}. 

\section{Machine-learning analysis}\label{sec:2}

The aim of this work is to examine the nature of 3FGL unassociated sources in order to select the best candidate sources, according to their predicted source class, for multiwavelength observations and to estimate the number of new $\gamma$-ray sources in each class that we might expect to identify in the future. Machine-learning algorithms are the best techniques for screening and classification of unassociated sources based on $\gamma$-ray data only. Such techniques were applied to 3FGL unassociated sources by \citet{mir16} to pinpoint potentially novel source classes, and by \citet{pablo} to classify them as likely AGN or PSR including, for the latter, predictions on the likely type of pulsar.

Focusing on the latter approach, they distinguished AGN from PSR 
using their $\gamma$-ray timing and spectral properties combining results from Random Forest \citep{bre01} and Boosted logistic regression \citep{fri00}. Out of 1010 unassociated sources, 559 were classified as likely AGN and 334 as likely PSR with an overall accuracy of $\sim$96 per cent. In addition, they used the same approach to classify pulsars into ``young'' and millisecond, leaving unexplored the distinction of AGN subclasses.

\begin{figure*}
\begin{center}
{\includegraphics[width=7.5cm, angle=0,clip=true]{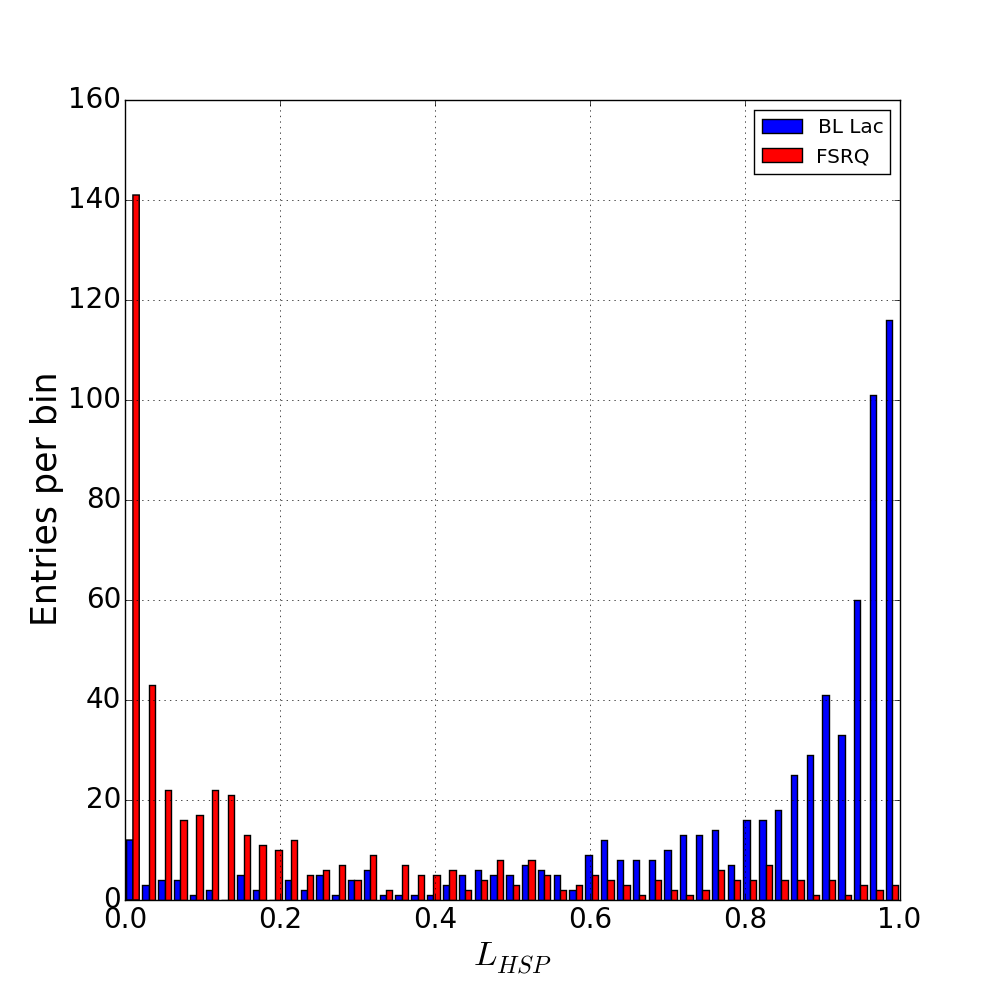}\hspace{.6cm}
\includegraphics[width=7.5cm, angle=0,clip=true]{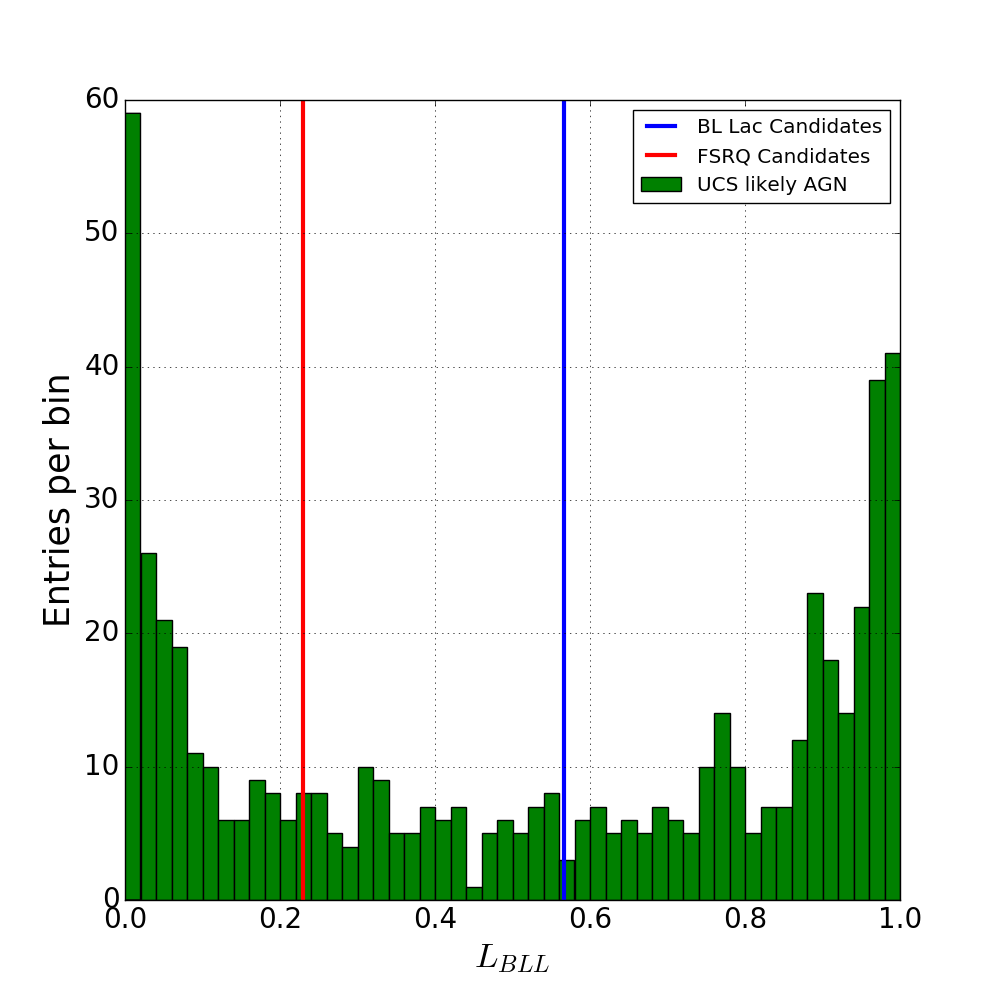}} 
\caption{(Left) Distribution of the ANN likelihood to be a BLL candidate for 3FGL BLLs (blue) and FSRQs (red). The distribution of the likelihood to be an FSRQ candidate ($L_{\textrm{\scriptsize FSRQ}}$) is 1--$L_{\textrm{\scriptsize BLL}}$. (Right) Same distribution for 559 3FGL unidentified sources classified as likely AGN by \citet{pablo}. Vertical blue and red lines indicate the classification thresholds of our ANN algorithm to label a source as BLL or FSRQ, respectively, as described in the text.}
\label{fig:ANN}
\end{center}
\end{figure*}

Here we want to integrate to the analysis performed in \citet{pablo} a classification of 559 3FGL unassociated sources likely AGN as likely BLL or FSRQ using the B-FlaP method described in \citet{bflap}. B-FlaP uses Empirical Cumulative Distribution Function (ECDF) and Artificial Neural Network (ANN) machine-learning techniques to classify blazars taking advantage of different $\gamma$-ray flaring activity for BLLs and FSRQs. We used a two-layer ANN algorithm \citep{bis95} to quantify the blazar flaring including as input the source parameters associated to 10 $\gamma$-ray flux values corresponding to the 10\%, 20\%, ..., 100\% fraction of observations below this flux. The output was set up to have two possibilities: FSRQ or BLL, with a likelihood ($L$) assigned to each so that  $L_{\textrm{\scriptsize BLL}}=1-L_{\textrm{\scriptsize FSRQ}}$. The closer to 1 is the value of $L$, the greater the likelihood that the source is in that specific source class. ANN was optimized using as source sample all 660 BLL and 484 FSRQ in the 3FGL catalogue through a learning method based on a standard back-propagation algorithm. The left-hand panel of Figure~\ref{fig:ANN} shows the likelihood distribution applied to all 3FGL blazars, which shows a clear separation of the two sub-classes of blazars based on flaring patterns.

Defining the \emph{precision} as the positive association rate, a $L_{\textrm{\scriptsize BLL}}$ value greater than 0.566 provides a \emph{precision} of 90 per cent for recognizing BLLs, while $L_{\textrm{\scriptsize BLL}}$ less than 0.230 identifies FSRQs with 90 per cent \emph{precision}.
Thanks to this approach, we have been able to apply B-FlaP to the full sample of 3FGL BCUs \citep{bflap}, obtaining statistical classifications for approximately 85 per cent of the sources\footnote{The list of B-FlaP BCU classifications is published online at:\\
\texttt{https://academic.oup.com/mnras/article/462/3/3180/2589794/Blazar-flaring-patterns-B-FlaP-classifying-blazar\#supplementary-data}}. Comparing the B-FlaP predictions with spectroscopic observations that were subsequently retrieved in literature \citep[e.g.][]{Vermeulen95, Titov11, Shaw13, AC16a, AC16b, Klindt17}, we note that there is a very good agreement between predictions and spectroscopic classifications. With 55 classified BCUs, 52 turn out to be spectroscopically confirmed, while 3, 2 FSRQ (3FGL J0343.3+3622, 3FGL J0904.3+4240) and one BLL (3FGL J1129.4$-$4215), are not consistent with the observations. While for 3FGL J0343.3+3622 ($L_{BLL} = 0.583$) and 3FGL J0904.3+4240 ($L_{BLL} = 0.673$) we have an intermediate likelihood value, meaning that the sources fall in a region characterised by a substantial overlap of the two different source classes, the case of 3FGL J1129.4$-$4215 is more problematic. With its low $L_{BLL} = 0.062$, it should be a high confidence FSRQ, while the observed counterpart definitely shows a BL Lac nature \citep{AC16b}. This source, however, has multiple possible counterparts (SUMSS J113014$-$421414 and SUMSS J113006$-$421441), all lying several arc minutes away from the signal centroid. In such circumstances, it may happen that the $\gamma$-ray signal is affected by contamination or badly associated, leading to the observed contradiction.

\citet{pablo} used any type of AGN to classify unassociated sources as likely AGN, including not only blazars, but also radio galaxies, compact steep spectrum quasars, Seyfert galaxies, narrow-line Seyfert 1s and other non-blazar active galaxies as well. These objects produce an overall contamination of $\sim$1.5 per cent to the blazar sample, slightly changing the value of \emph{precision} of the ANN given the classification thresholds. As a result, the \emph{precision} for recognizing BLLs and FSRQ decreases to 87 per cent, introducing a contamination given by non-blazars of $\sim$2 per cent for the former and $\sim$ 3 per cent for the latter.

\section {Results and validation}\label{sec:3}

In this section we discuss the results of our optimized ANN algorithm at classifying BLL and FSRQ candidates  among 3FGL unassociated sources. Applying our optimized algorithm to the 559  unassociated sources classified as likely AGN by \citet{pablo}, we find that 271 are classified as BLL candidates, 185 as FSRQ candidates, and 103 remain likely AGN of uncertain type. The right-hand panel of Figure~\ref{fig:ANN} shows the likelihood distribution applied to likely AGN, which reflects very well those of known BLL and FSRQ. Interestingly, we find that the ratio of likely BLL to FSRQ obtained by our analysis ($\sim$ 1.4) is very similar to the ratio of known BLL and FSRQ (1.4).

Table~\ref{tab:3} shows a portion of individual results of the classification of 3FGL unassociated sources classified as likely AGN, where, for each source, we provide the ANN likelihood ($L$) to be a BLL or an FSRQ, and the predicted classification according to the defined classification thresholds. The second and  third columns of the list show the Galactic longitude and latitude respectively. The full table is available electronically from the journal.

\begin{table*}
\caption{Classification list of 3FGL unassociated sources classified as likely AGN. The columns are: the name in the 3FGL catalogue, the Galactic longitude and latitude ($b$ and $l$), the ANN likelihood to be classified as a BLL ({$L_{\textrm{\scriptsize BLL}}$}) and a FSRQ ({$L_{\textrm{\scriptsize FSRQ}}$}) and the predicted classification. The full table is available online.}\label{tab:3}
\begin{center}
\begin{tabular}{lcccccc}
\hline
\hline
{\bf 3FGL Name}	&	{\bf l ($^{\circ}$)}	&	{\bf b ($^{\circ}$)}	& {\bf \bll } 	& {\bf \fsrq} 	&	{\bf Classification}	\\
\hline
J0000.2--3738	&	345.41	&	--74.947	&	0.986	&	0.014	&	BLL	\\
J0001.6+3535	&	111.66	&	--26.188	&	0.842	&	0.158	&	BLL	\\
J0002.0-6722	&	310.14	&	--49.062	&	0.974	&	0.026	&	BLL	\\
J0003.5+5721	&	116.49	&	--4.912	&	0.712	&	0.288	&	BLL	\\
J0004.2+0843	&	103.6	&	--52.363	&	0.903	&	0.098	&	BLL	\\
J0006.2+0135	&	100.4	&	--59.297	&	0.772	&	0.228	&	BLL	\\
J0006.6+4618	&	114.91	&	--15.867	&	0.383	&	0.617	&	AGN Uncertain	\\
J0007.4+1742	&	108.33	&	--43.911	&	0.825	&	0.175	&	BLL	\\
J0007.9+4006	&	113.98	&	--22.007	&	0.856	&	0.144	&	BLL	\\
J0014.3--0455	&	99.59	&	--66.096	&	0.118	&	0.882	&	FSRQ	\\
\hline
\end{tabular}
\end{center}
\end{table*}

\begin{figure*}
\begin{center}
{\includegraphics[width=7.5cm, angle=0,clip=true]{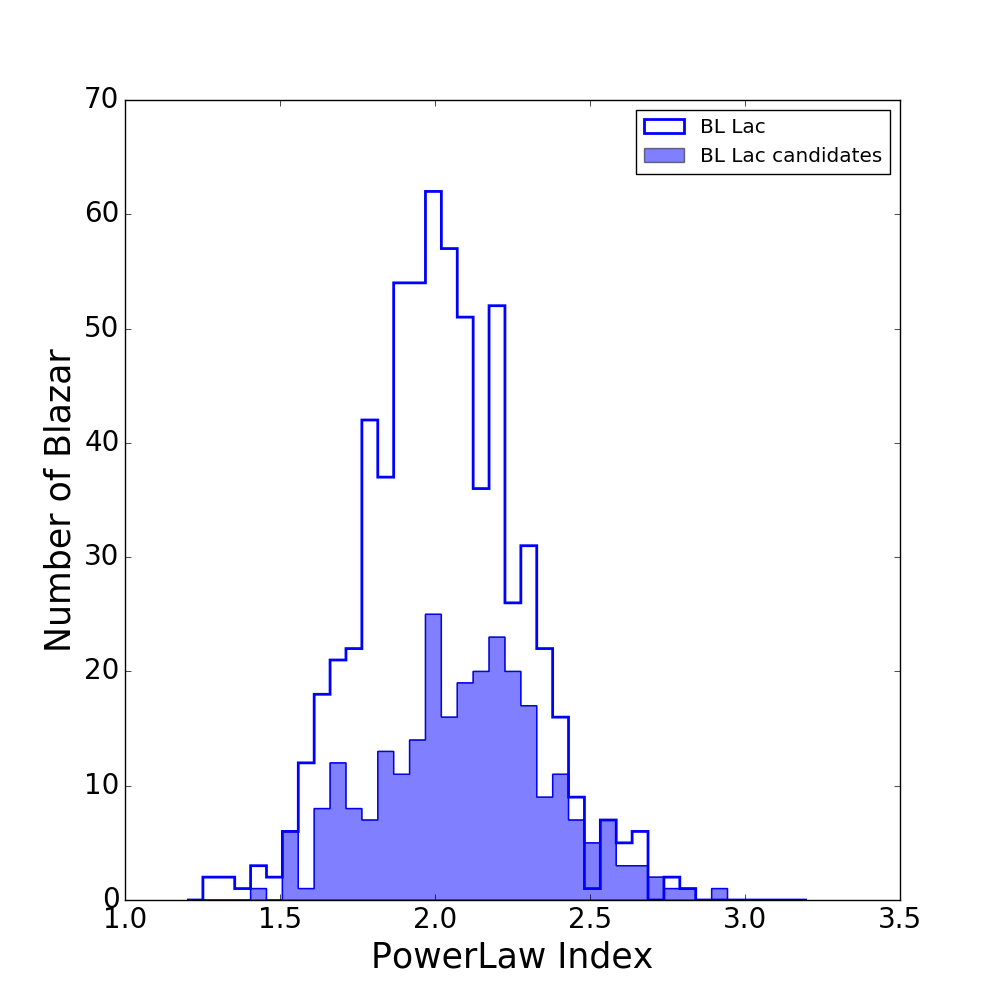}\hspace{.6cm}
\includegraphics[width=7.5cm, angle=0,clip=true]{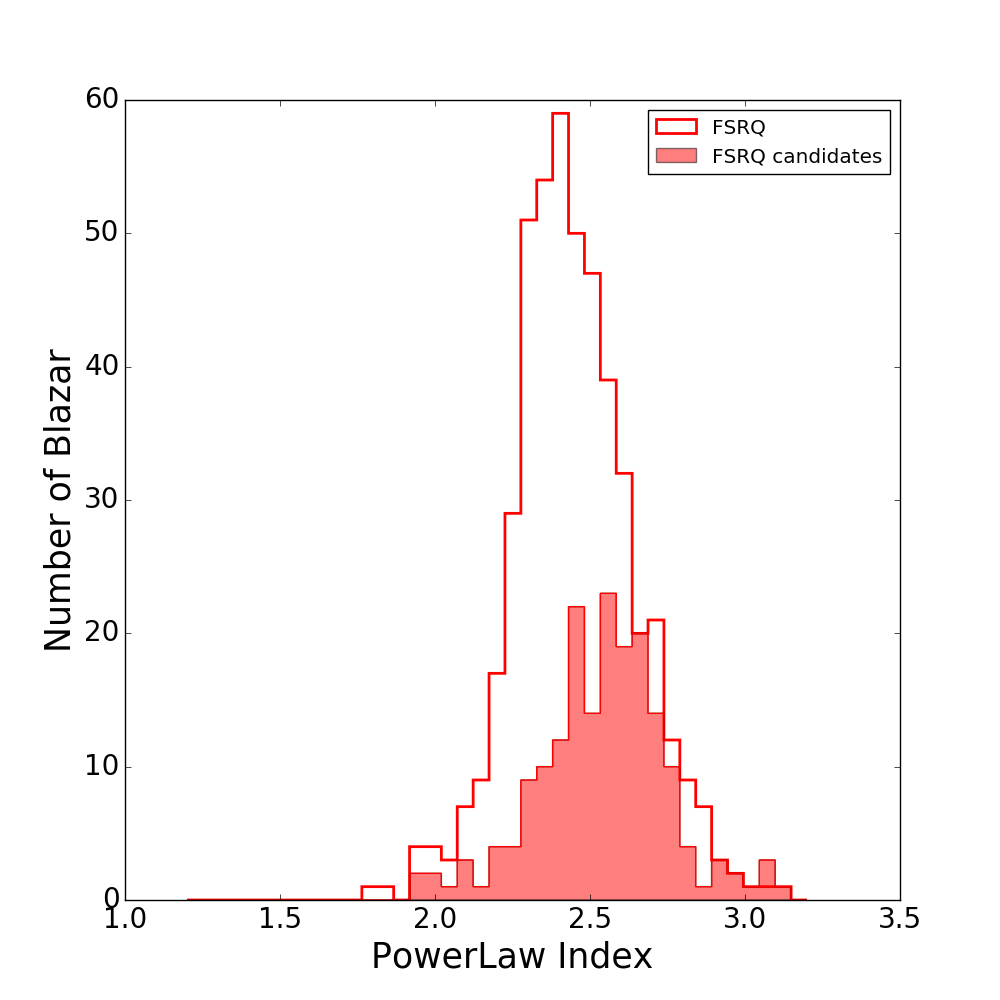}} 
\caption{Power-law index distribution for the unidentified sources classified as blazars by the ANN method (filled histograms) in comparison to the previously classified blazars. Left: BLL; right: FSRQs.}\label{fig:powlaw}
\label{plaw}
\end{center}
\end{figure*}

Since we did not include any spectral information in the ANN algorithm, we validate our results comparing the spectra for known BLL and FSRQ with those of likely blazar subclasses. Gamma-ray BLL have average spectra that are flatter than those of FSRQs \citep{ack15}. The best-fitting photon spectral
index (in 3FGL named power-law index) distribution has a mean value of 2.02$\pm$0.25 for the former and 2.45$\pm$0.20 for the latter, where uncertainties are reported at the 1$\sigma$ confidence level. Figure~\ref{fig:powlaw} shows that the power-law index distribution for likely BLL and FSRQ is consistent with those of known BLL and FSRQs (mean value of 2.10$\pm$0.27 for the former and 2.54$\pm$0.21 for the latter).

Another way to validate the predictions of our method is to compare them with classifications obtained after the release of the 3FL catalogue. Currently, optical spectroscopic observations campaigns to hunt blazars among unassociated $\gamma$-ray sources are ongoing \citep[see e.g.][]{lan15,mas16,alv16,mar16}. These follow-up multiwavelength classification efforts have resulted in 24 new blazar associations, 21 classified as BLL and 3 as FSRQs. Since our algorithm was optimized to select the best targets to observe in other wavelengths, we can evaluate the performance of our method analyzing the positive association rate (\textit{precision}). Out of 24 new blazars with optical spectra, B-FlaP classifies 22 as BLL, while 2 remain unclassified. For the subset of 22 BLL candidates, our prediction matches in about 90 per cent of the objects with optical spectra, in agreement with our classification thresholds definition, while we cannot assert anything about the \textit{precision} in the classification of FSRQ. This result give a strong confirmation about the optimal performance of our classification algorithm even if the number of new blazar associations is still small.

\subsection{Optical spectroscopic observations}\label{sec:4}

Encouraged by optimal performance of our classification algorithm, we carried out optical spectral observations at the Asiago Astrophysical Observatory of the best targets within the unassociated source sample classified as likely AGN. The new observations were executed with the 1.82m {\it Copernico} telescope\footnote{Website: \url{http://www.oapd.inaf.it/index.php/en}} and the 1.22m {\it Galileo} telescope\footnote{Website: \url{http://www.dfa.unipd.it/index.php?id=300}}, configured for long-slit optical spectroscopy, with the instrumental configurations reported in Table~\ref{tabAsiago}. Both telescopes are able to provide moderate spectral resolution data ($R \sim 600$) over a wavelength range spanning 3700\AA\ to 7500\AA, achieving a continuum signal to noise ratio (SNR) of order $\sim 20$ in 2 hours of exposure on targets with an optical magnitude $V \sim 17$.

\begin{table*}
  \caption{Instrumental characteristics of the Asiago Astrophysical Observatory Telescopes. \label{tabAsiago}}
  \begin{center}
    \begin{tabular}{lcc}
      \hline
      \hline
      {\bf Instrumental characteristic} & {\bf Copernico Telescope} & {\bf Galileo Telescope}\\
      \hline
      Main mirror diameter & $1.82\,$m & $1.22\,$m \\
      Focal length & $5.39\,$m & $6.00\,$m \\
      Spectrograph & AFOSC$^{\rm a}$ & Boller \& Chivens  \\
      Entrance slit width & $1.69\arcsec$  & $3.5 - 5.0\arcsec$\\
      Grating & $300\,$gr mm$^{-1}$ & $300\,$gr mm$^{-1}$\\
      Wavelength range & $3700$--$8000\,$\AA & $3500$--$7500\,$\AA \\
      Spectral resolution & $600$ & $600$ \\
      \hline
    \end{tabular}
  \end{center}
\begin{flushleft}\hspace{3.5cm}
      {\footnotesize
        $^{\rm a}$Asiago Faint Object Spectrograph and Camera
        }
\end{flushleft}
\end{table*}

Taking into account the observational limitations introduced by the magnitude constraints and the geographical position of the observing site, we performed an observing campaign, selecting the targets for observations from the list of 3FGL {unassociated sources}. Due to the quite large uncertainties on the positions of $\gamma$-ray sources, especially faint ones, the identification of the plausible optical counterparts to be observed was obtained by combining radio and X-ray observations, in a similar way to the method described in \citet{bflap}. The reason for combination lies in the theoretical expectation that a high-energy source, powered by a jet of relativistic charged particles, which produce $\gamma$-ray photons, should suffer significant energy losses from synchrotron radiation at radio and X-ray frequencies \citep*{Boettcher12}. Since the spatial resolution of detectors operating at such lower energies is far better than in the $\gamma$-ray band (down to a few arcseconds), the optical counterparts were associated with objects emitting both radio and X-ray photons, within the $\gamma$-ray signal confinement area at the 95 per cent confidence level. Although this technique proved reliable to support the association of BCU targets to optical counterparts, in the case of unassociated sources, further care was required. These sources are generally faint and with quite large positional uncertainties. Consequently, also the expected synchrotron losses are weak and might be missed in standard X-ray and radio surveys. In general we are able to select the low-energy candidate counterpart by matching the NRAO VLA Sky Survey \citep[NVSS]{Condon98} with the {\it Swift} satellite X-ray catalogue \citep[1SXPS]{Evans14} in circular regions of 10$'$ in radius, centered on the $\gamma$-ray signal centroid. In some cases, however, the sources are too weak to be listed in a catalogue, particularly at X-rays, and a subsequent reanalysis of the observations of the target is the only way to pinpoint the most likely optical counterpart to such sources.

\begin{figure*}
\includegraphics[width=0.48\textwidth]{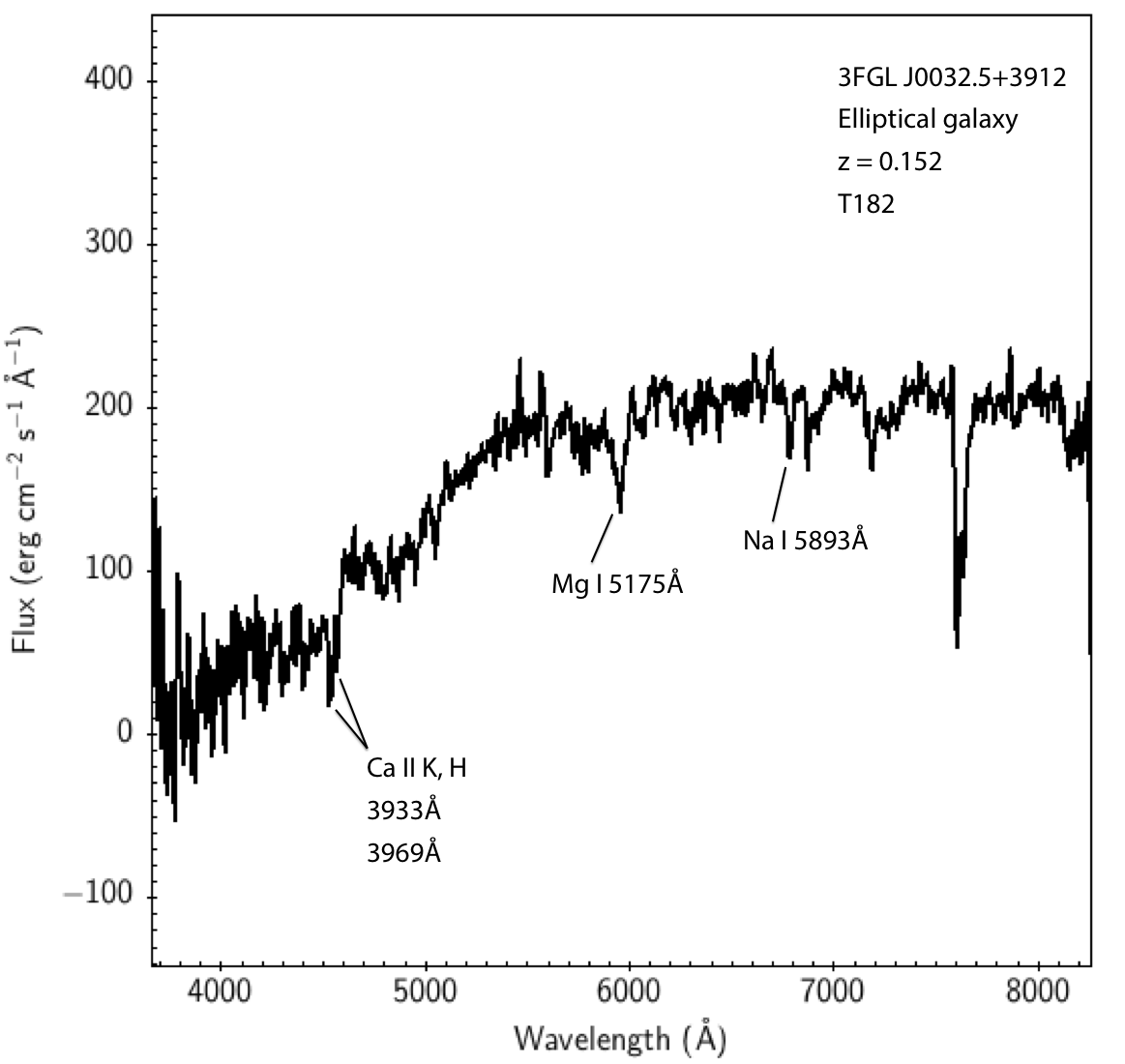}
\includegraphics[width=0.48\textwidth]{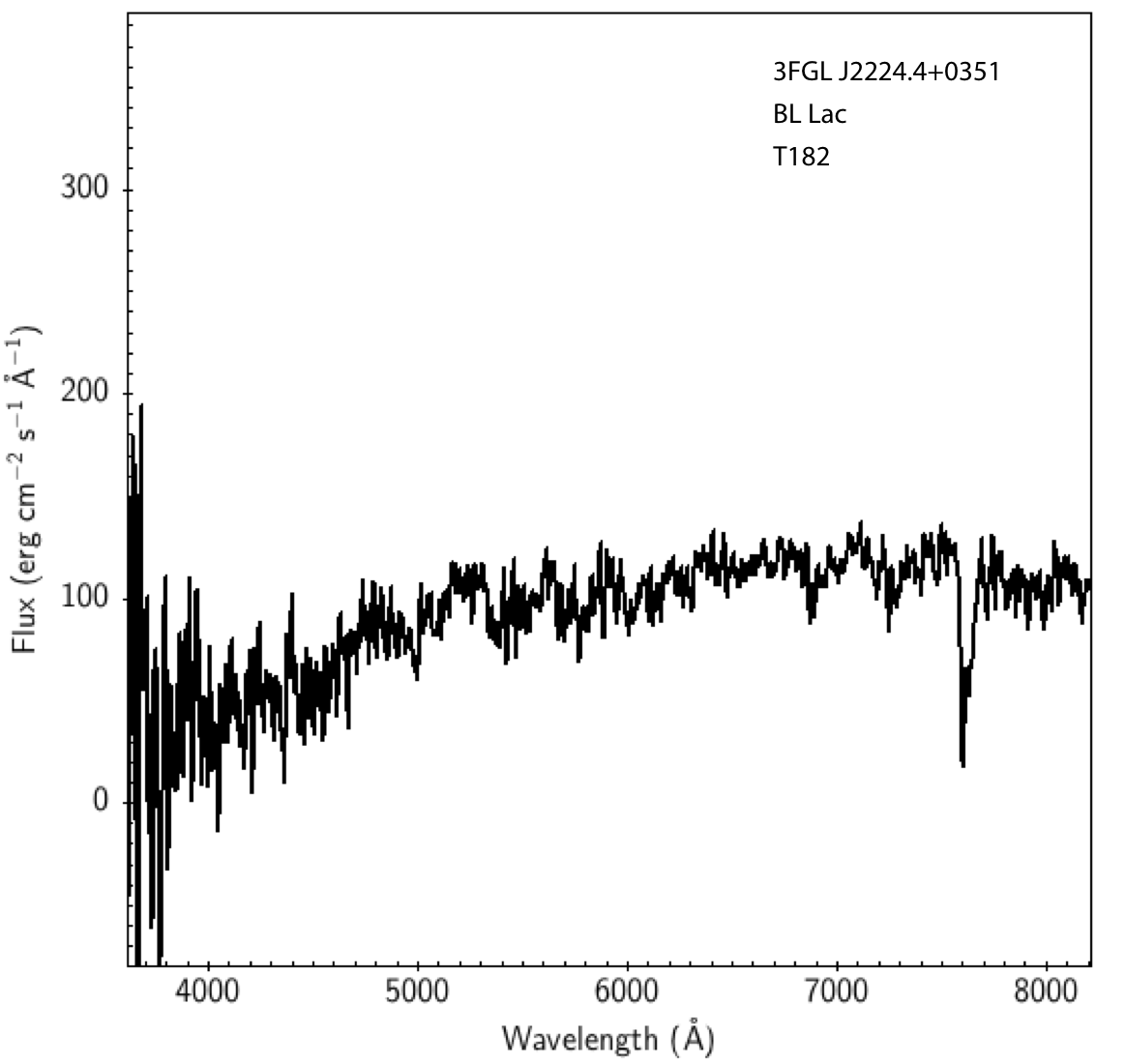}
\includegraphics[width=0.48\textwidth]{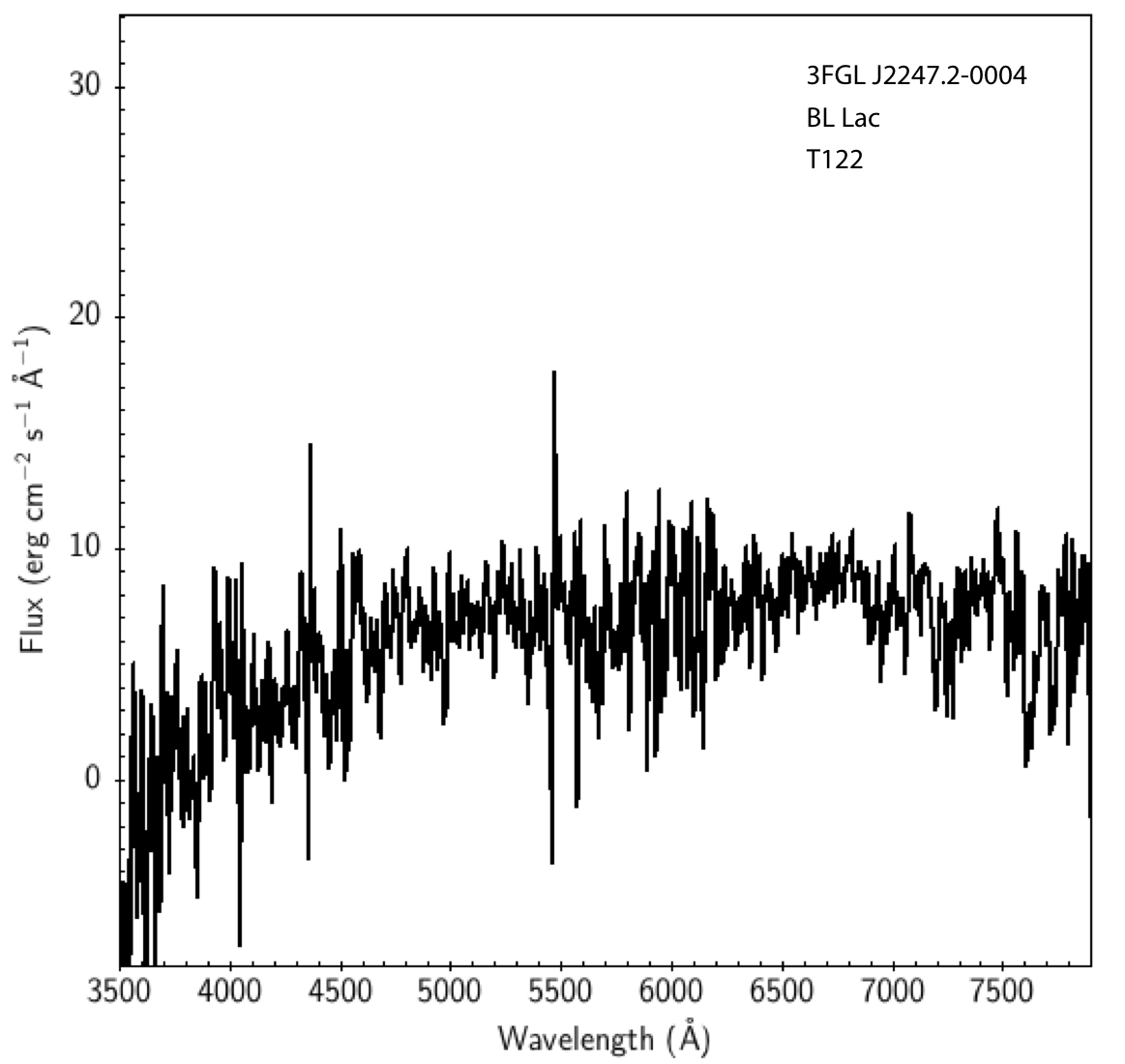}
\includegraphics[width=0.48\textwidth]{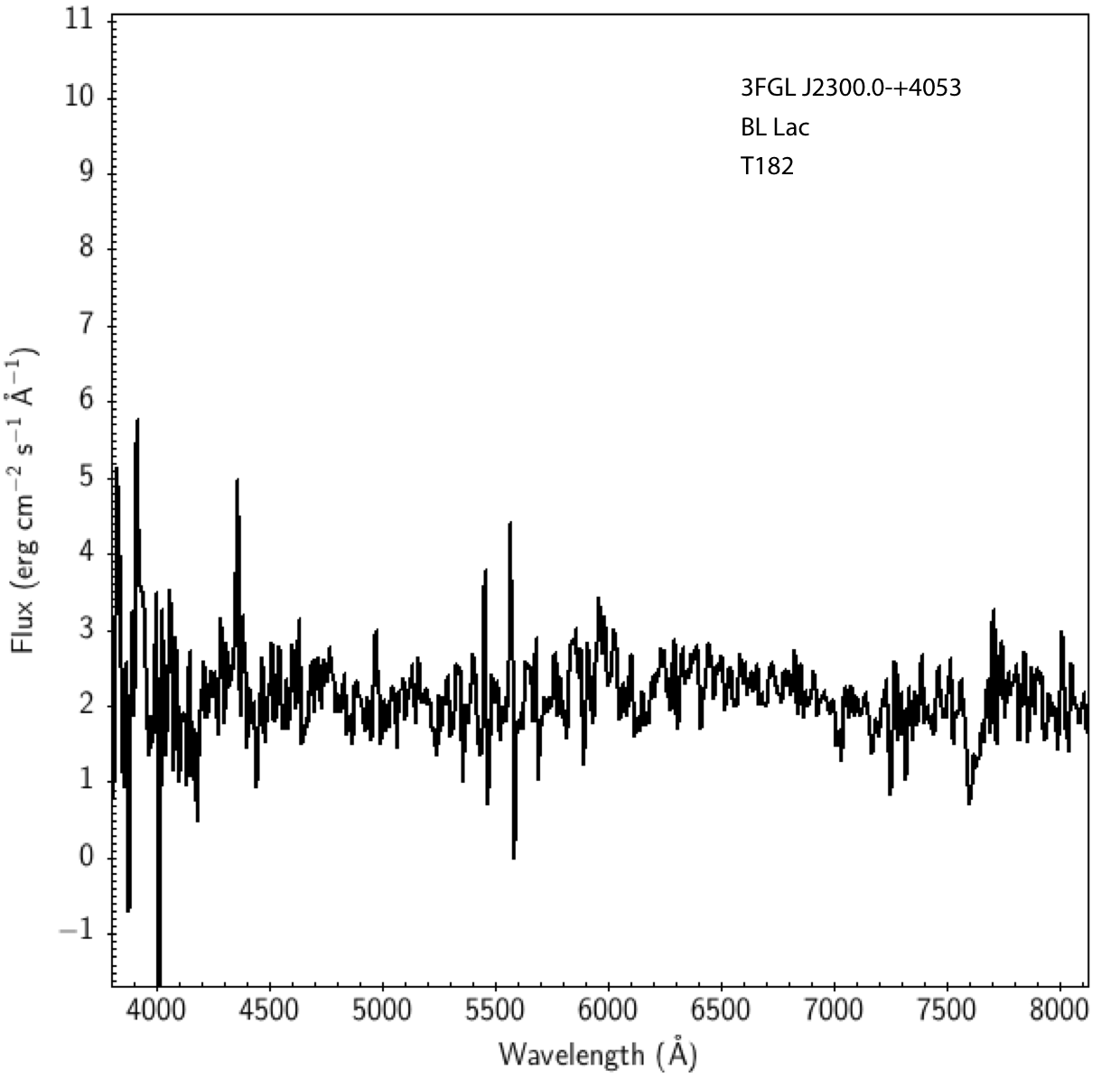}
\includegraphics[width=0.48\textwidth]{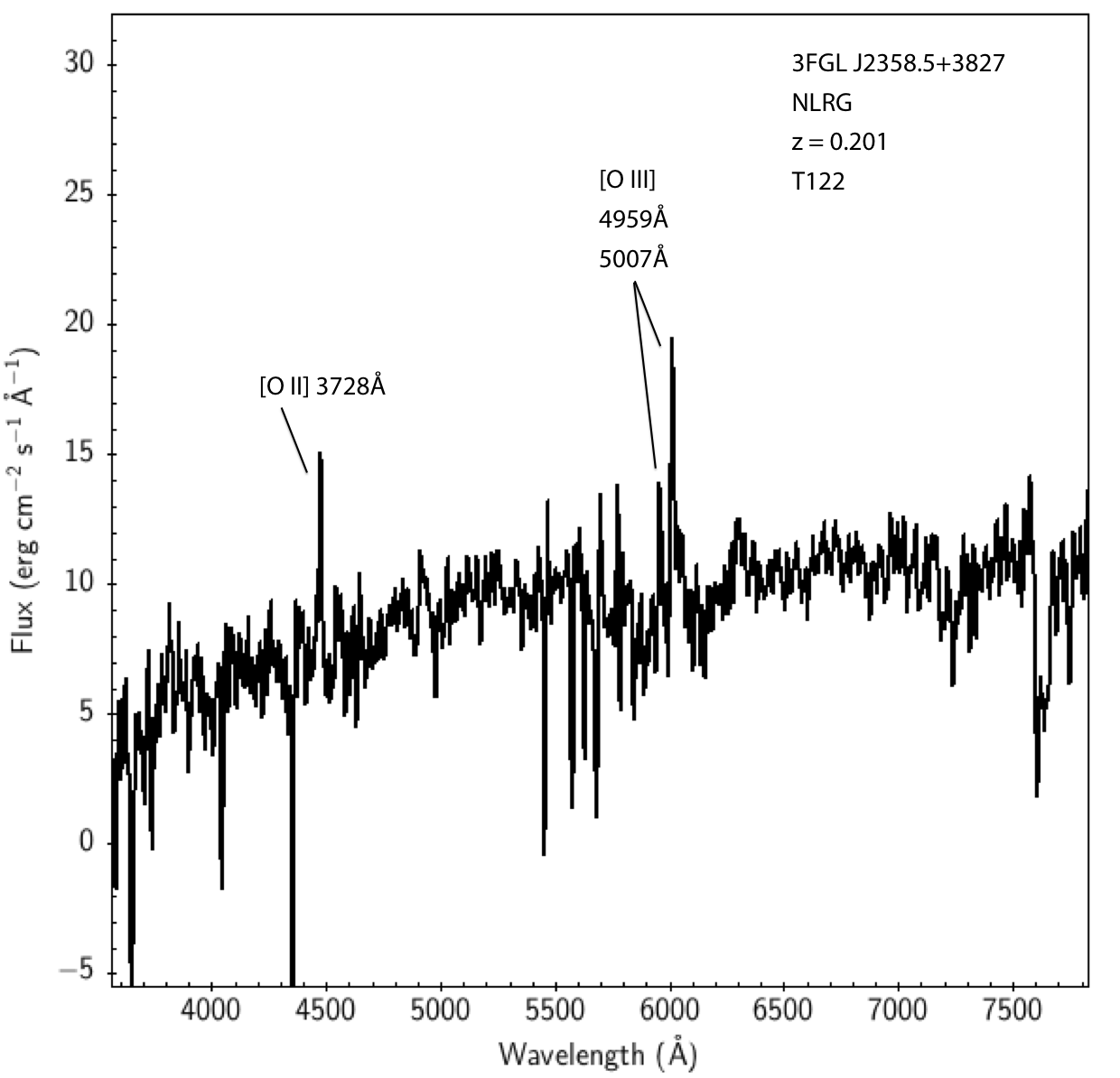}
\caption{Optical spectra of 3FGL unassociated sources observed from the Asiago Astrophysical Observatory. \label{figAsiago}}
\end{figure*}

When all the observational constraints were satisfied, we observed the targets, collecting a total of 2 hours of exposure time for every target, split in observations lasting from 20 up to 30 minutes each. The exact duration of the single exposures was determined by the best trade-off between the requirement to improve the SNR, the need to track the spectrum on the surface of the detectors, and the contamination from cosmic rays and night sky emission lines. These background contributions, indeed, may lead to saturation effects on the detector, with consequent loss of spectral information, and must therefore be removed. Combining several short exposures to form a longer observation can immediately filter out the cosmic ray background, since it follows a random pattern that is easily identified and masked out from the complete data set. The same process leads to an efficient subtraction of the sky emission lines, because the combined spectrum is not subject to saturation limits and  can collect an arbitrarily high signal, in order to interpolate the sky contribution from regions close to the source and to remove them from the spectrum. All the observations were taken together with comparison FeAr arc lamp spectra to perform wavelength calibration, and they were paired with observations of spectro-photometric standard stars, to provide flux calibration. We proceeded with the standard long slit spectroscopic data reduction procedure, which involves bias subtraction and flat field correction, by means of standard IRAF tasks\footnote{Website \url{http://www.iraf.noao.edu}}, arranged in a specific pipeline that is optimized to work with the Asiago telescopes instrumental configuration. In this study, we obtained 5 new spectra,  illustrated in Fig.~\ref{figAsiago}. We detail in the following the characteristics of these spectra and the resulting insights.

\subsubsection{3FGL 0032.5+3912}
This object has been associated with an optical source that, once observed with the 1.82m telescope, showed the characteristic spectrum of an elliptical galaxy. The identification of absorption lines like the Ca~{\small II} $\lambda\lambda$3933,3969 doublet (detected at 4530\AA\ and 4572\AA), together with Mg~{\small I} $\lambda$5175 and Na~{\small I} $\lambda$5893 (detected at 5961\AA\ and 6789\AA) places the redshift of this object at $z = 0.152$. Its optical spectrum is consistent with an elliptical galaxy that may host BLL activity, in agreement with the BLL classification suggested by the machine learning technique.

\subsubsection{3FGL 2224.4+0351}
Observed with the 1.82m telescope, this object shows a featureless continuum spectrum, with a peak close to 7000\AA, subsequently decaying towards the short wavelength regions. No clear signs of emission and absorption lines are detected in the spectrum, consistent with the predicted BLL classification of the source.

\subsubsection{3FGL 2247.2-0004}
This source has been observed with the 1.22m telescope and shows a featureless continuum spectrum, decaying towards the short wavelength regime, much like the previous case. The lack of clear emission and absorption lines is consistent with its predicted BLL classification.

\subsubsection{3FGL 2300.0+4053}
The spectrum of this source shows a more prominent power-law continuum that increases towards the short wavelength regime. With the exception of some unidentified spike-like features, very likely descending from increase of noise due to lower detector efficiency, the classification turns out to be consistent with a source of BLL type.

\subsubsection{3FGL 2358.5+3827}
When observed with the 1.22m telescope, this source shows a clear system of emission lines that can be identified as the close [O~{\small II}] $\lambda\lambda$3727,3929 doublet (detected as an unresolved feature at 4477\AA) and the strong [O~{\small III}] $\lambda\lambda$4959,5007 doublet (falling at 5956\AA\ and 6013\AA), which place this source at redshift $z = 0.201$. The detection of strong narrow lines, together with an underlying continuum that becomes weaker at short wavelength, are suggestive of an obscured AGN activity. The source has a 1.4GHz flux measured by the NVSS of
$$F_{1.4\,{\rm GHz}} = 57.4 \pm 1.8\,{\rm mJy}$$
that, adopting a standard $\Lambda$ Cold Dark Matter Cosmology with $H_0 = 70\,{\rm km\, s^{-1}\, Mpc^{-1}}$, $\Omega_\Lambda = 0.7$ and $\Omega_M = 0.3$, corresponds to a distance of 920.3 Mpc and to an intrinsic luminosity
$$\nu L_\nu = (5.49 \pm 0.17) \cdot 10^{40}\,{\rm erg\, cm^{-2}\, s^{-1}}.$$
This suggests that this object could be classified as a Narrow Line Radio Galaxy (NLRG), in spite of the predicted classification as a BLL, although the detection of such type of objects in $\gamma$-rays, at the inferred redshift, is extremely rare.

\section{Discussion and conclusions}\label{sec:5}

One of the main goals of our investigation is to complete the census of blazar subclasses in the 3FGL source catalogue using the ANN technique based on B-FlaP \citep{bflap}. B-FlaP is well suited to perform preliminary and reliable classification of likely blazars when detailed observational or multiwavelength data are not yet available. This is the typical situation for almost all unassociated sources in \emph{Fermi}-LAT catalogues. Recently, \citet{pablo} applied a number of machine-learning techniques to classify 3FGL unassociated sources as likely pulsar or AGN, focusing only on the former, to identify the most promising unassociated source to target in pulsar search. 

We applied our algorithm to 559 3FGL unassociated sources classified as likely AGN to investigate their source subclass. These sources can be divided in 271 BLL candidates, 185 FSRQ candidates, leaving only 103 without a clear classification. We validated our predictions comparing their $\gamma$-ray spectra with the expected ones. In addition, we compared our results with the source classes inferred by recently published optical spectroscopic observations \citep{lan15,mas16,alv16,mar16}. This comparison results in 29 new blazar associations, out of which 5 are obtained thanks to our new optical observations. For the subset of 27 overlapping sources, our prediction matches in $\sim$ 90 per cent of the objects as expected. Such excellent agreement confirms the power of our method as a classifier for unidentified sources as well. Our work can help to identify targets both for blazar searches and for follow-up studies of blazars at very-high $\gamma$-ray energies with ground-based imaging air Cherenkov telescopes (MAGIC, HESS, VERITAS).

\citet{lef17} have recently published a paper aimed at researching blazar candidates among the {\it Fermi}-LAT 3FGL catalogue using a combination of boosted classification trees and multilayer perceptron artificial neural networks methods. Their work is divided in two steps. In the first one they applied the combined classifier to separate 3FGL unassociated sources as blazar or pulsar candidates, while in the second one they use the same approach to determine the BLL or FSRQ nature of both blazar candidates and of BCU. In contrast to our approach, they used both spectral and timing $\gamma$-ray parameters to separate source classes. Out of 595 blazar candidates among the 3FGL unassociated sources, \citet{lef17} study the blazar subclass nature for 417 sources that have no caution flag as described in \citet{3FGL}. Out of these, 371 match with our blazar candidate sample. Applying their classifier to this sample they divide them into 192 BLL and 129 FSRQ candidates. The comparison with our corresponding subset of 223 BLL candidates shows that our prediction is in agreement with \citet{lef17} for 174 objects (about 80 per cent) and in disagreement for 28 (about 12 per cent). We observe that 13 objects in disagreement are characterized by a very low prediction value (L$_{BLL}<0.7$), thus the discrepancy between the two approaches decreases significantly when defining a more robust classification threshold. In addition, comparing our subset of 83 FSRQ candidates we observe that our prediction is in agreement for 62 sources (about 75 per cent) and in disagreement for 8 (about 9 per cent). Interestingly, analysing the PowerLaw Index distribution for the sources that are in disagreement with \citet{lef17} we observe that they are located in the region of overlap between BLL and FSRQ (between 2.1 and 2.9) making difficult their classification including even spectral information. Only future optical spectroscopic observations will unveil the real nature of these sources. As a result, although \citet{lef17} applied a very different approach from our work, we find a good overall agreement, indicating that both methods are useful classifiers. We obtained the same result applying our optimised algorithm to all 417 un-flagged sources classified as blazar candidates by \citet{lef17}.

\begin{figure}
\resizebox{\hsize}{!}{\includegraphics{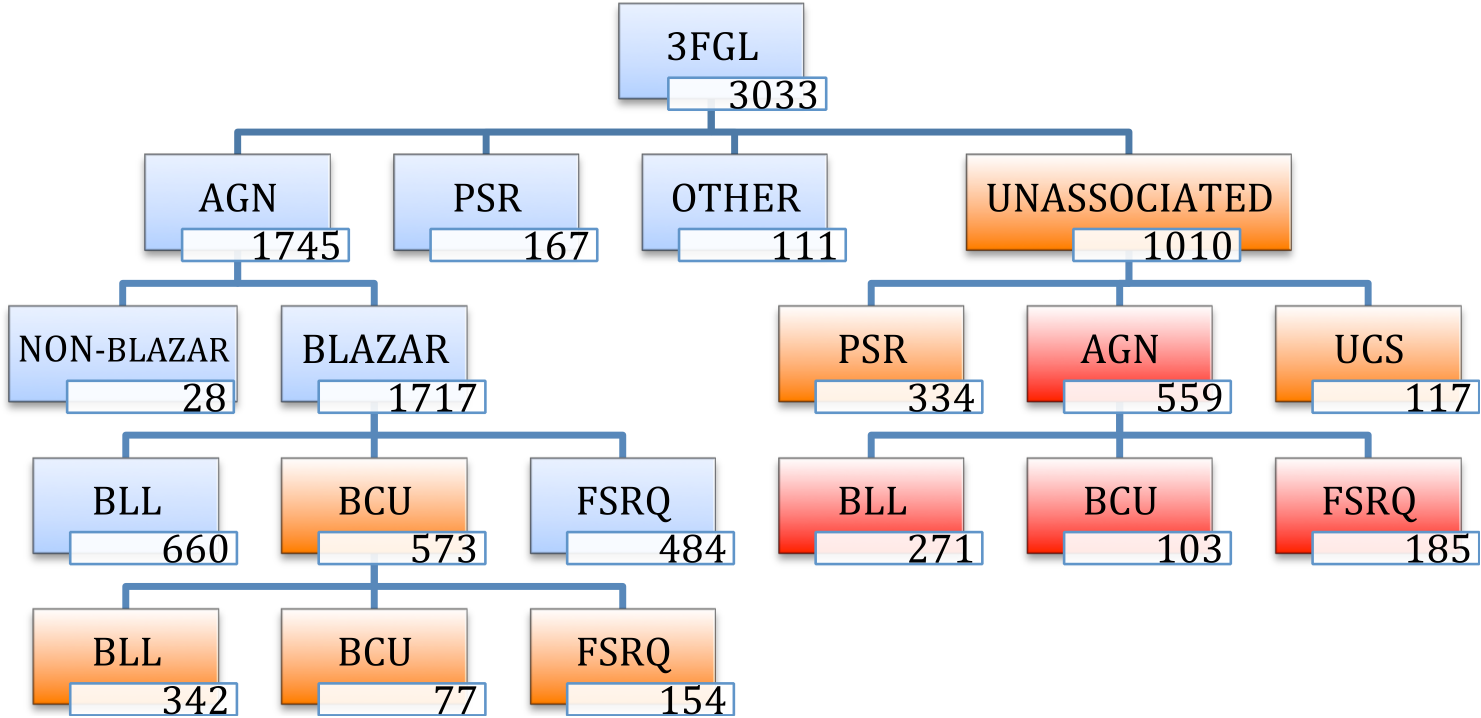}}
\caption{3FGLzoo after this work. Orange: source classification provided by \citet{bflap} and \citet{pablo}. Red: our classification of unassociated sources classified as likely AGN. Here UCS are unassociated sources that are not classified as PSR or AGN candidates.}\label{fig:6}
\end{figure}

Putting together the overall result of this study with the ones obtained in \citet{bflap} and \citet{pablo} we can characterize the entire $\gamma$-ray
population proposing a new distribution of 3FGL sources, as shown in Figure~\ref{fig:6}, where cells in red represent results obtained in this work. Table~\ref{tab:4} shows the number of $\gamma$-ray sources per each class reported in the 3FGL catalogue and after this work. The number of BLL (or candidates) increased by a factor of 1.9, while that of FSRQ of 1.7, raising the ratio of BLL to FSRQ from 1.36 to 1.55. Interestingly, out of 180 blazars of uncertain type, only 20 (11 per cent) are located at low Galactic latitude (|b|<10 deg). We expect that a very small fraction (less than 3 per cent) of non-blazar AGN subclasses (Seyferts, radio galaxies and other AGN) could contaminate the sample of blazar of uncertain type. As an important result, the efforts aimed at classifying 3FGL sources decreased the fraction of uncertain sources (BCU and unassociated sources) from 52 per cent to 10 per cent, discriminating the best targets for future  follow-up multi-wavelength observations.

\begin{table}
\caption{The new classification of the most numerous $\gamma$-ray sources classes combining this study with \citet{bflap} and \citet{pablo} in comparison with the 3FGL catalogue.}\label{tab:4}
\begin{center}
\begin{footnotesize}
\begin{tabular}{lccccccr}
\hline
\hline
\bf{Class} & \bf{3FGL} & \bf{Post 3FGL} \\
\hline
Blazar & 1717 (57\%) & 2276 (75\%)\\
\quad -- BLL & 660 & 1273 \\
\quad -- FSRQ & 484 & 823 \\
\quad -- BCU &573 & 180 \\
Pulsar &167 (6\%) & {\bf 501 (17\%)}\\
Unassociated & 1010 (33\%) & {\bf 117 (4\%)}\\
Others & 139 (4\%) & 139 (4\%)\\
\hline
\end{tabular}
\end{footnotesize}
\end{center}
\end{table}

\section*{Acknowledgements}
We thank the anonymous referee for his/her very helpful comments and suggestions to our manuscript. Support for science analysis during the operations phase is gratefully acknowledged from the \emph{Fermi}-LAT collaboration for making the 3FGL results available in such a useful form, the Institute of Space Astrophysics and Cosmic Physics of Milano -Italy (IASF INAF) and The Goddard Space Flight Center NASA. DS acknowledges support through EXTraS, funded from the European Commission Seventh Framework Programme (FP7/2007-2013) under grant agreement n. 607452. We thank Pablo Saz Parkinson because this paper  builds directly on his work. This work is based on observations collected at Copernico telescope (Asiago, Italy) of the INAF - Osservatorio Astronomico di Padova and on observations collected with the 1.22m \textit{Galileo} telescope of the Asiago Astrophysical Observatory, operated by the Department of Physics and Astronomy ``G. Galilei'' of the University of Padova.

\bsp	
\label{lastpage}
\end{document}